\documentclass{pasj01}
\usepackage{mathrsfs} 
\usepackage{graphicx} 
\usepackage{mediabb}
\usepackage{color}  
  
\title{Giant Cometary \textsc{Hii} Regions and Molecular Bow Shocks in Spiral Arms of Galaxies: M83}
\author{Yoshiaki \textsc{Sofue}\altaffilmark{} }
\altaffiltext{}{Institute of Astronomy, The University of Tokyo, Mitaka, Tokyo 181-0015} 
\email{sofue@ioa.s.u-tokyo.ac.jp}

\begin{document} 
\date{ } 
\maketitle   

 \def\Msun{M_\odot} \def\Lsun{L_\odot} 
 \def\deg{^\circ} \def\r{\bibitem[]{}} \def\/{\over}\def\kms{km s$^{-1}$}  
  \def\sin{{\rm sin}\ } \def\cos{{\rm cos}\ } \def\Hcc{ H cm$^{-3}$ }  
  \def\co{$^{12}$CO$(J=1-0)$ } 
\def\be{\begin{equation}} \def\ee{\end{equation}} 
\def\Te{T_{\rm e}}\def\ne{n_{\rm e}}\def\ue{u_{\rm e}}  
\def\mH{m_{\rm H}} \def\Hcc{H cm$^{-3}$} 
\def\Halpha{ H$\alpha$ }\def\nuv{N_{\rm UV}}
\def\ne{n_{\rm e}}\def\ni{n_{\rm i}}\def\nh{n_{\rm H}}
\def\ar{\alpha_{\rm r}} \def\L{L_{\rm UV}}
 
\def\Rbow{R_{\rm bow} } \def\Rhii{R_{\rm HII}}    \def\Rzero{R_0}    
\def\Lcone{L_{\rm cone} } \def\Acone{\Theta }

\def\Hii{\textsc{Hii} }
\def\Tn{T_{\rm n} } \def\Te{T_{\rm e} } \def\Ha{H$\alpha$ }
\def\({\left(} \def\){\right)} \def\[{\left[} \def\]{\right]}    
\def\Kkms{ K \kms } \def\Hsqcm{ H cm$^{-2}$ }
\def\Ico{I_{\rm CO} }
   
\def\tc{t_{\rm col}} \def\rhoc{\rho_{\rm cloud} } \def\tb{ t_{\rm bow} }
\def\rhom{\rho_{\rm mean} } \def\Rc{R_{\rm cloud} } 
\def\V{ V_{\rm rot} } \def\Vp{ V_{\rm p} } \def\Op{ \Omega_{\rm p} }
\begin{abstract} 
A number of giant cometary HII regions (cones) (GCH) sheathed inside molecular bow shocks (MBS) are found along spiral arms of the barred galaxy M83. The open cone structure is explained by a model of expanded HII front in a gaseous arm with steep density gradient, and the bow shock is shown to be formed by encounter of an HII region with the supersonic gas flow across the arm. It is suggested that dual-side compression of molecular gas at the bow head between the MBS and GCH enhances star formation along the spiral arms.
\end{abstract}   

\section{Introduction}
 
Astrophysical bow shock is a classical subject, and is observed around objects interacting with supersonic gas flows in a wide range of scales from planets to cosmic jets (Dyson et al. 1975; van Buren et al. 1990; Ogura et al. 1995; Wilkin 1996;  Arce and Goodman 2002;  Reipurth et al. 2002). Galactic-disk scale bow structure is observed in spiral arms, where a supersonic flow in galactic rotation encounters a stagnated gaseous arm in the density-wave potential (Martos and Cox 1998; G{\'o}mez and Cox 2004a, b). 

\begin{figure} 
\begin{center}      
\includegraphics[width=7cm]{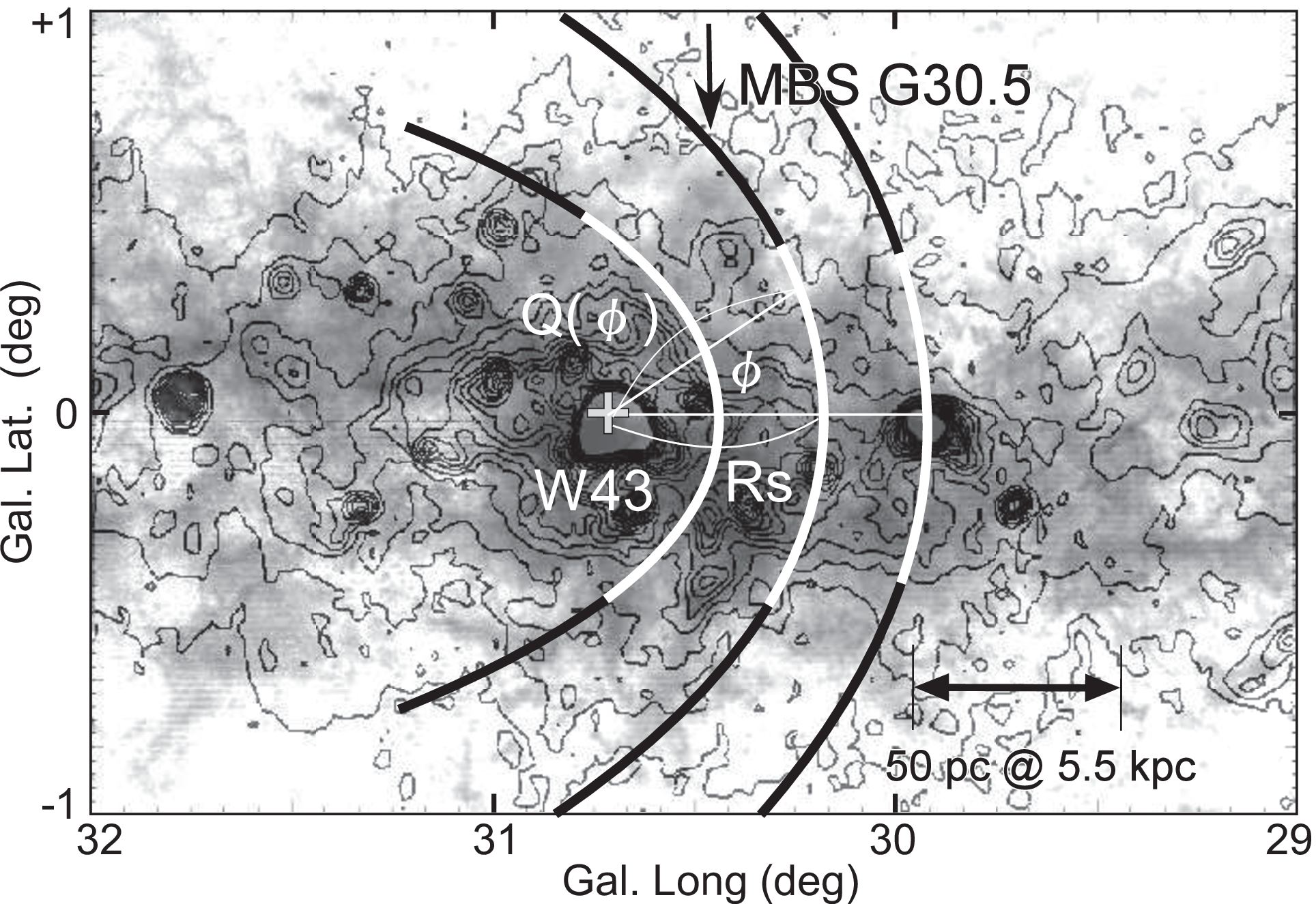}  
\end{center}
\caption{\co intensity map (gray) of the molecular bow shock (MBS) G30.5 in the Galaxy overlaid on a 10-GHz continuum map (contours) (Sofue et al. 2018)), which compose concave arc with respect to W43. Thick lines indicate calculated bows for $\Rbow=25$, 50 and 75 pc.}
\label{G30}
\end{figure}

In scales of star-forming (SF) regions, a bow shock was observed at G30.5+00 associated with the SF region W43 in the tangential direction of the 4-kpc molecular arm (Scutum arm) in thermal radio continuum emission associated with a CO line molecular arc (Sofue 1985). The molecular bow at G30.5+00 has recently been studied in detail based on the Nobeyama 45-m CO line survey (Sofue et al. 2018), which we call the molecular bow shock (MBS) (figure \ref{G30}). 

\begin{figure} 
\begin{center}      
\includegraphics[width=7.5cm]{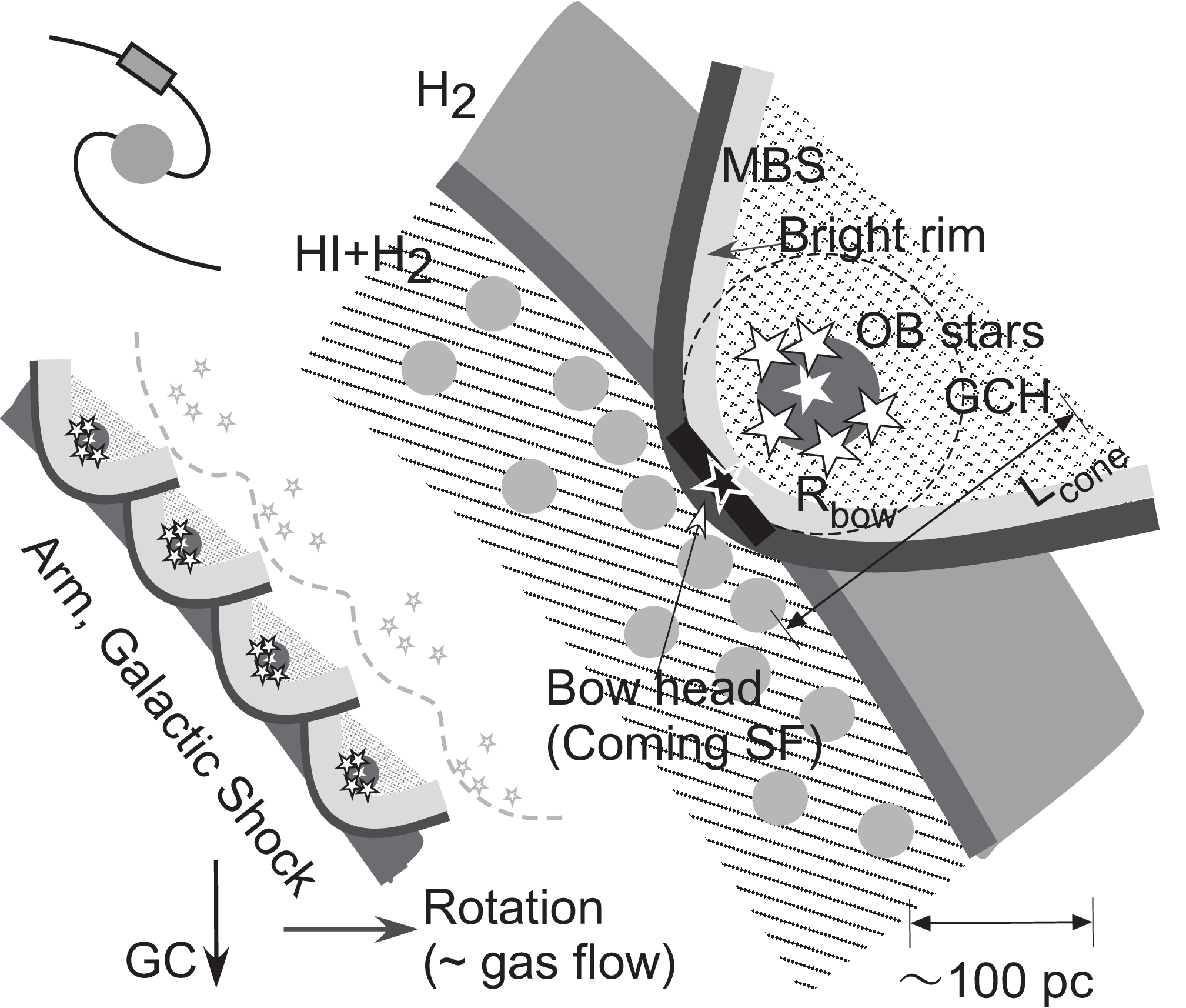}   
\end{center}
\caption{Illustration of the molecular bow and GCH concave to the central OB star cluster proposed for the W43 SF complex in the 4-kpc arm in the Galaxy (Sofue et al. 2018). Inserted small is an illustration for a wavy sequential star formation discussed later.  }
\label{illust}
	\end{figure}

MBS is a concave arc of molecular gas around an HII region (SF region) formed in the up-stream side of galactic rotation with respect to the SF region. An MBS is formed in such a way that the interstellar gas in galactic supersonic flow encounters a pre-existing HII region on the down stream side in the galactic-shock wave (figue \ref{illust}).

A similar phenomenon is observed in star forming regions known as a cometary HII region tailing down stream, when a compact HII region is embedded in a flow of ambient interstellar gas (Arthur \& Hoare 2006; Reid \& Ho 1985; Steggles et al. 2017; van Buren et al. 1990; Fukuda and Hanawa 2000; Campbell-White et al. 2018; Deharveng et al. 2015). The current studies of cometary HII regions have been obtained of sub-parsec to parsec scale objects inside individual SF regions. However, our observations of the association of the spiral-arm scale MBS G30.5 and the SF complex around W43 suggests the existence of larger scale, spiral-arm scale cometary HII regions associated with MBS. Namely, MBS developed in the up-stream side of cometary HII regions may be a common phenemenon in spiral arms.

On such premise, we have searched for bow-shock plus cometary structure in spiral arms of nearby galaxies. In the present paper, we report identification of a number of bow structures in the barred spiral galaxy M83 (NGC 5236). Assuming that dark clouds in optical images represent molecular clouds, we name them  molecular bow shocks (MBS), We also show that MBS are generally associated with giant cometary HII regions (GCH) on their down-stream sides, which may alternatively be called giant HII cone (GHC). Therefore, an MBS and a GCH make one single set of objects. So, they may be often referred to either MBS or GCH.

Morphology and energetics (luminosity) of individual HII regions and OB clusters have been studied by optical imaging of M83 using the Hubble Space Telescope (HST) (Chandar et al. 2010, 2014; Liu et al. 2013;  Blair et al. 2014;  Whitmore et al. 2011). High-resolution molecular gas distribution in M83 has been extensively observed in the CO line emissions, and detailed comparative study with HII regions are obtained using ALMA high resolution maps (Hirota et al. 2018; Egusa et al. 2018).  

Structural relation of HII regions and molecular clouds has been one of the major subjects of star formation mechanism in the Galaxy such as cloud-cloud collisions (McKee and Ostriker 2007). However, spatially-resolved relation between indiviidual HII regions and dark clouds in external galaxies seems to have not been studied yet. We here focus on individual HII regions and morphological relation with their associated dark clouds in M83. We will show that their morphology is similar to the cometary cone structure modeled for G30.5 in the Galaxy. In order to explain the morphology, we propose qualitative models based on theories of bow shocks and expanded HII regions.

\section{Extragalactic Giant Cometary \Hii Regions (GCH) and Molecular Bow Shocks (MBS)} 
        
\begin{figure*} 
\begin{center}       
\includegraphics[width=16cm]{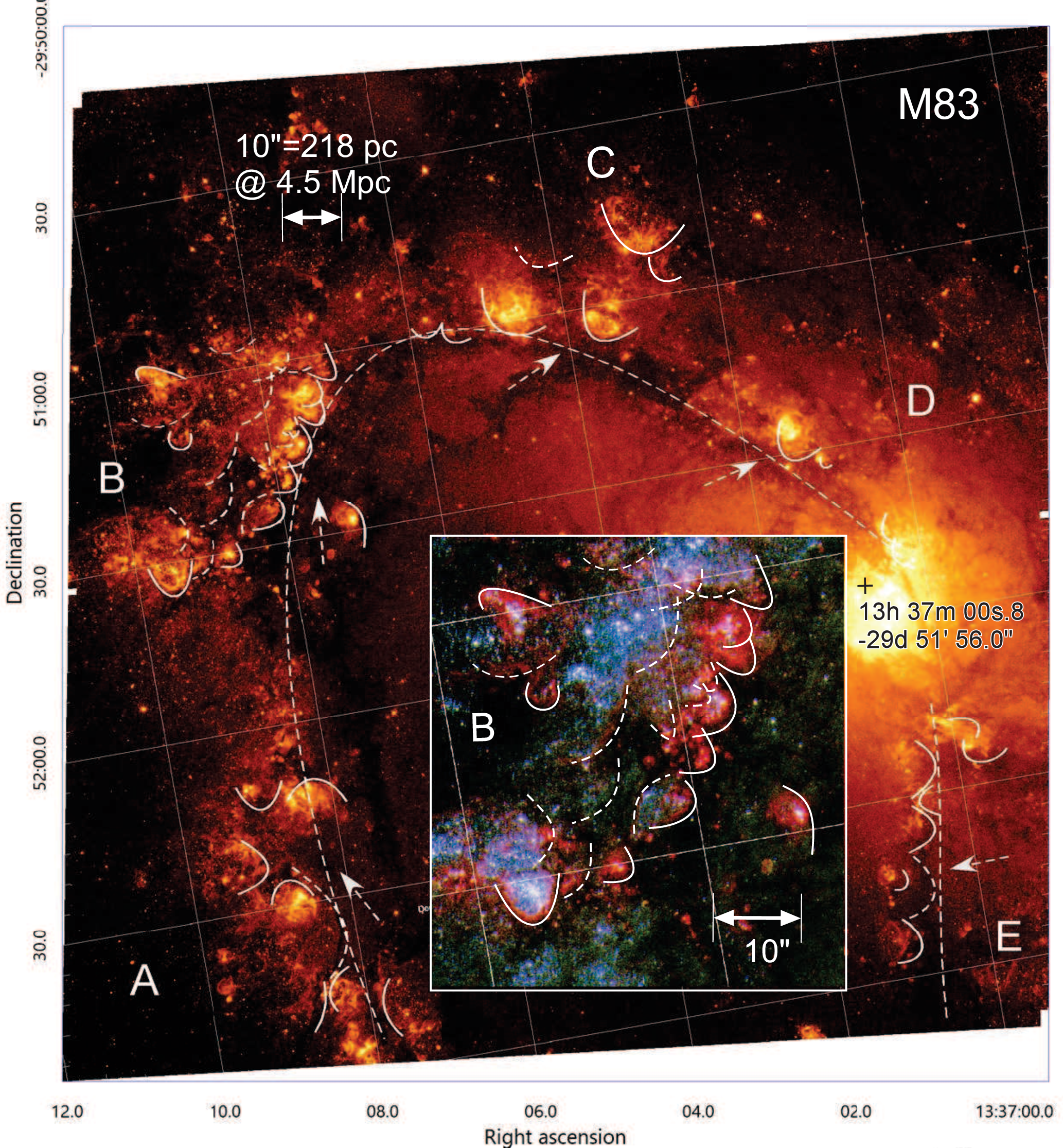}  
\end{center}
\caption{Giant cometary HII regions (GCH) and molecular (dark) bow shocks (MBS) in M83 marked by white arcs on a $\lambda 657$-nm band image taken with the HST (http://www.stsci.edu/hst/wfc3/phot-zp-lbn). Spiral arm and rotation directions are indicated by a dashed line and arrows, respectively. Inserted is an enlargement of region B in RGB (657, 502, 438 nm) color composite. } 
\label{zoo}
	\end{figure*}

\begin{figure*} 
\begin{center}       
\includegraphics[width=15cm]{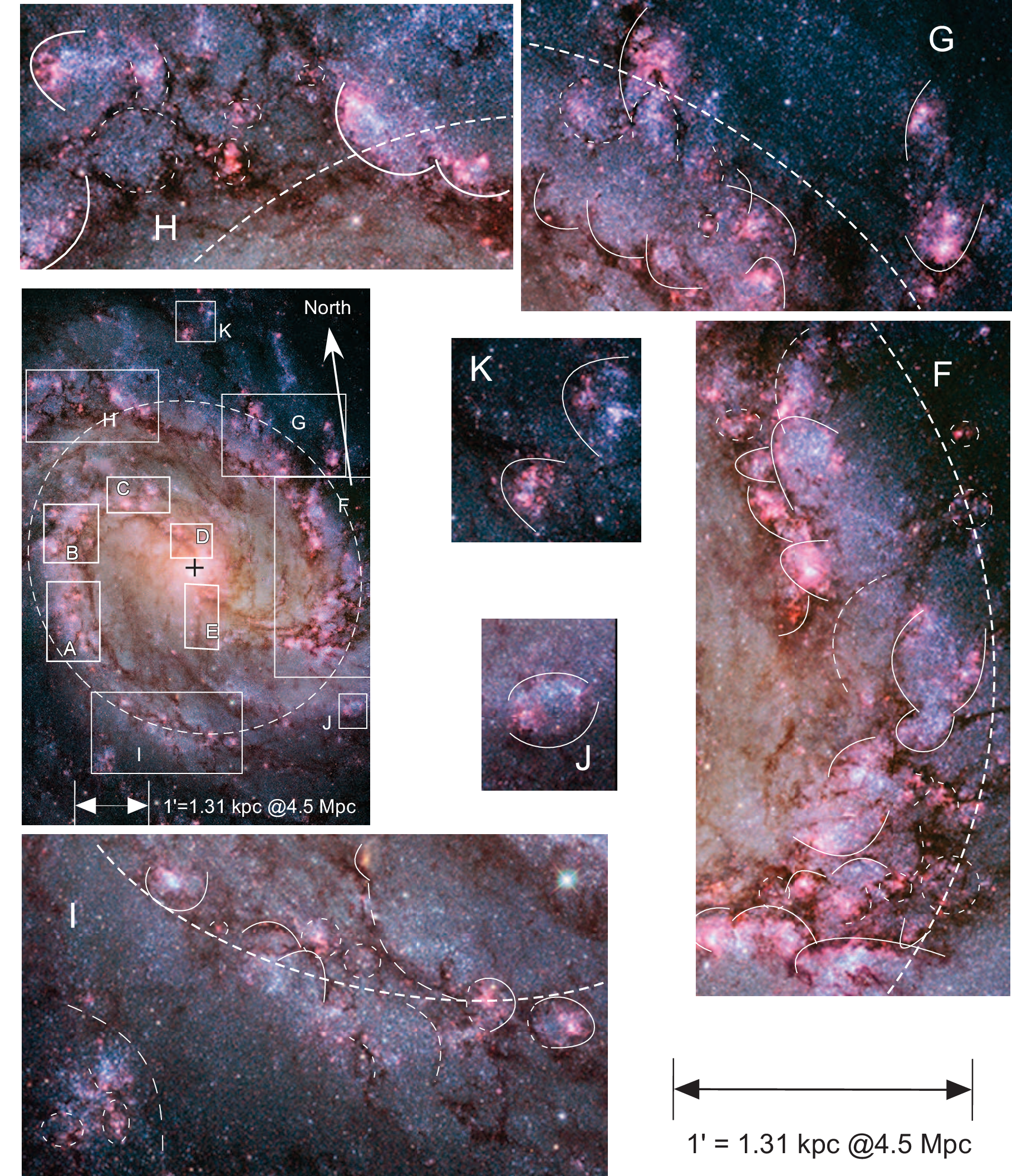}  
\end{center}
\caption{Same as figure \ref{zoo}, but for outer regions of M83. The dashed large ellipse and long dashed lines indicate corotation circle (Hirota et al. 2014). Thin dashed arcs and small ellipses are less clear or irregular MBS+GCH.  Color photo was taken from the web page of NASA at https://apod.nasa.gov/apod/ap140128.html.} 
\label{zoo-II}
	\end{figure*}

We examine optical images of M83 in $\lambda$438, 502 and 657 nm bands observed with the Hubble Space Telescope (HST) taken from the STScI and NASA web sites.
We adopt the nucleus position at RA=13h 37m 00s.8  and Dec=$-29\deg 51' 56''.0$ (Sofue and Wakamatsu 1994) and a distance of 4.5 Mpc (Thim et al. 2003). 
Figure \ref{zoo} shows an HST image of M83 at 657 nm, where are marked giant cometary HII regions (GCH) associated with dark bow-shaped features (MBS) by white arcs. Figure \ref{zoo-II} shows the same, but as supplement for the outer region.
 
The identification of MBS+GCH was obtained by eye-identification as follows, while numerical identification might be desirable, but is far beyond the scope of the present skills in imaging astronomy. First, it was easy to find HII regions and open clusters using each of the three color photographs (e.g., red for HII regions and blue for OB clusters), as well as assisted by color-coded image as inserted in figure \ref{zoo}.  Then, a search for associated dark lanes and clouds was obtained, and young, therefore bright SF/HII regions are almost allways associated with dark clouds. We excluded too faint or diffuse HII regions, which are either associated with diffuse clouds or not associated with dark cloud.

By pairing an HII region with a dark cloud, their morphological relation was looked into in detail individually. In most cases, an HII region is surrounded by an arc of dark lane in such a way that the HII region is lopsided and open to the interarm direction, while the other, brigher side is facing a concave bow-shaped dark lane, as illustrated in figure \ref{illust}. Thus found lopsided HII region and a molecular arc are here identified as a GCH (giant cometary HII region) and MBS (molecular bow shock), assuming that a dark cloud is a molecular cloud.

The GCH and MBS are generally located on the down-stream sides of dark lanes of spiral arms. Each GCH is sheathed inside an MBS, and the inner wall of MBS coincides tith the outer front of the HII region composing a bright rim of \Ha emission. Figure \ref{illust} illustrates the GCH/MBS strucutre.

HII regions in M83 have been classified into several categories according to the sizes and luminosities (Whitmore et al. 2011). We here classify HII regions in M83 into three morphological types, and focus on Type III, and are summarized in table \ref{tabclass}.
\begin{itemize}
\item Type I: HII bubbles around low luminosity OB clusters with sizes smaller than the disk thickness and galactic shock thickness with full extent less than $\sim 30$ pc. HII regions of categories 1 to 3 by Whitmore et al. (2011) are of this type.
\item Type II: Bipolar cylindrical HII region open to the halo with length comparable to the disk scale height. The wall looks like a hole in the disk with diameter comparable to the disk thickness.  Categories 3 to 4 are of this type.
\item Type III: Giant cometary HII regions (GCH), alternatively giant HII cones (GHC), open to the halo as well as to inter-arm, which develop around luminous OB associations. The extent is comparable to or greater than the disk thickness and the width of galactic shock wave with full extent as large as $\sim 100-200$ pc. This type of HII regions are of categories 4 to 6. 
GCH is characterized by the curvature $\Rbow\sim \Rzero$ of the bow/cometary head, axis length $\Lcone$, and opening angle of the cone. 
\end{itemize}

	\begin{table*}  
\caption{Morphological classification of HII regions.\\}
        \begin{center}
\begin{tabular}{llllll}  
\hline 
\hline 
Type & property&$\Rbow$ (pc) & $\Lcone$ (pc)  &$\L(\Lsun)$\\   
\hline 
I & HII bubble& $\sim 10$  &$ \sim 20$ &$\sim 10-10^2$\\
II& Bipolar HII cylinder &$\sim 20$ &$\sim 20-50$ &$\sim 10^2$\\
III & Giant cometary HII cone &$\sim 20-150$ &$\sim 50-200$ & $\sim 10^2-10^5$ \\ 
\hline  
\end{tabular}
\end{center}
\label{tabclass}  
	\end{table*}

Using figure \ref{zoo}, where the coordinates are indicated, we measured the positions, bow-head curvatures $\Rbow$, and position angles (PA) of the GCH/MBS by fitting the arcs by parabola. The measured values are listed in table \ref{tabbow}.

\begin{figure} 
\begin{center}        
\includegraphics[width=6cm]{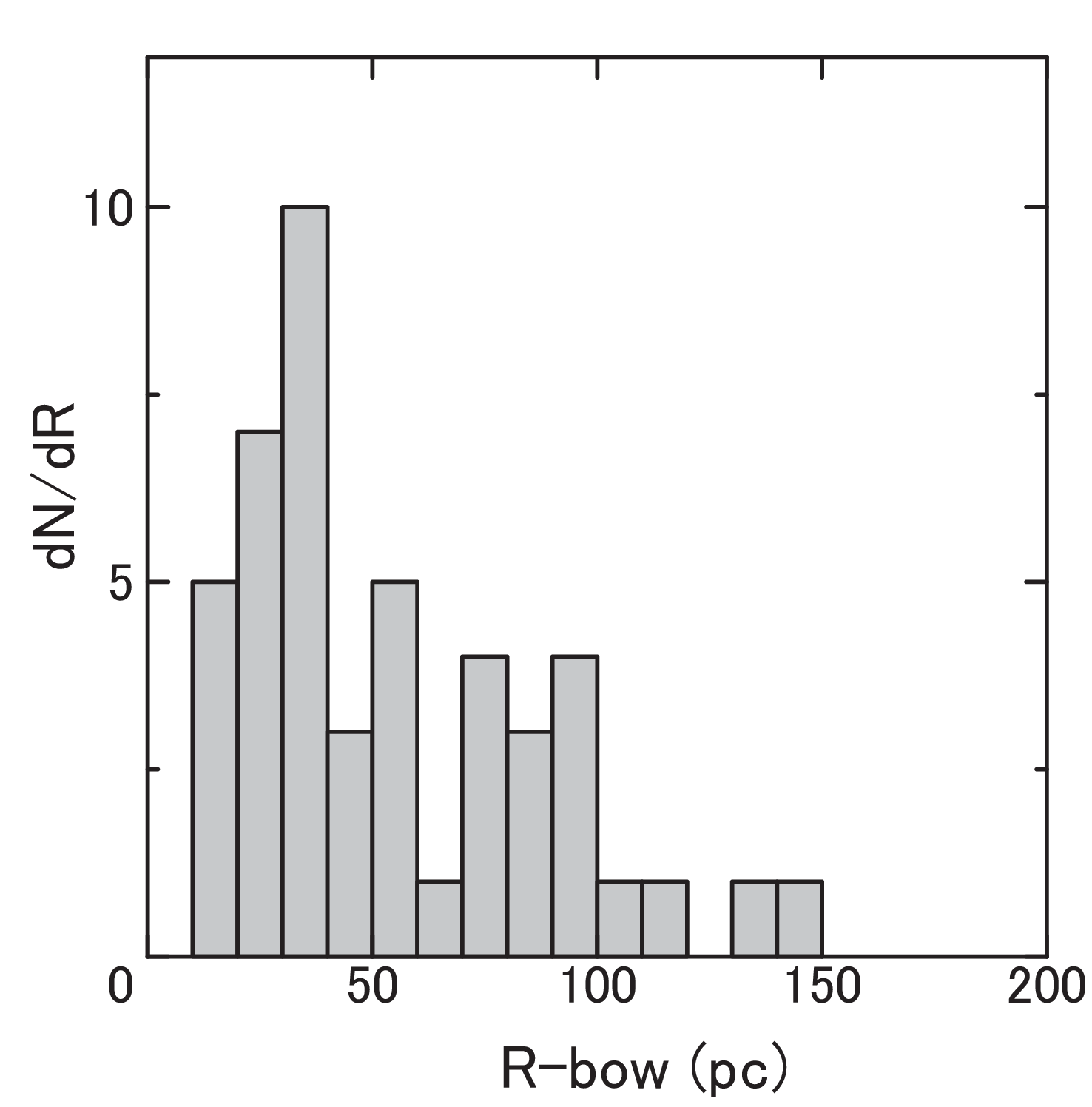}\\   
\includegraphics[width=6cm]{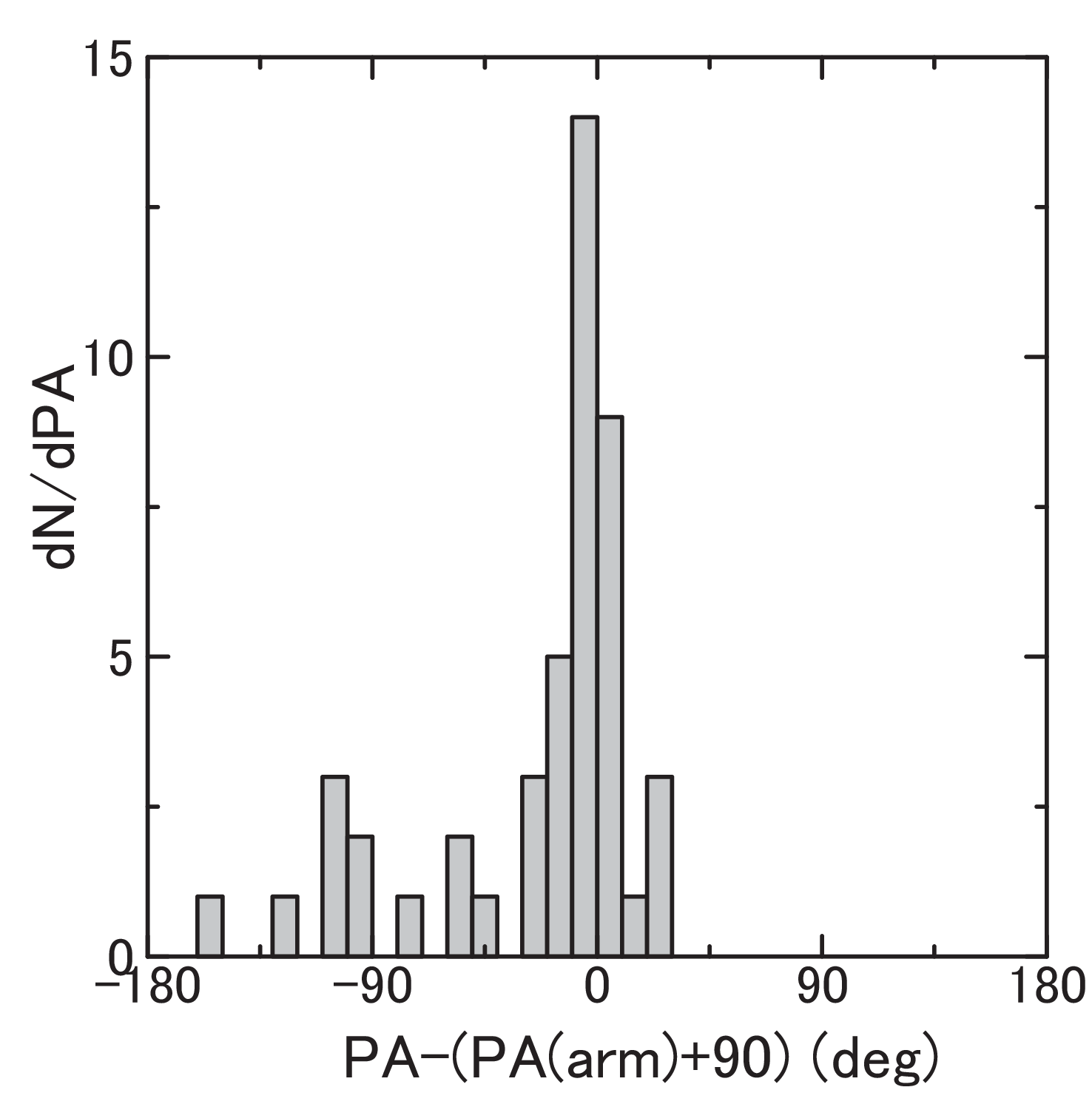}\\  
\includegraphics[width=6cm]{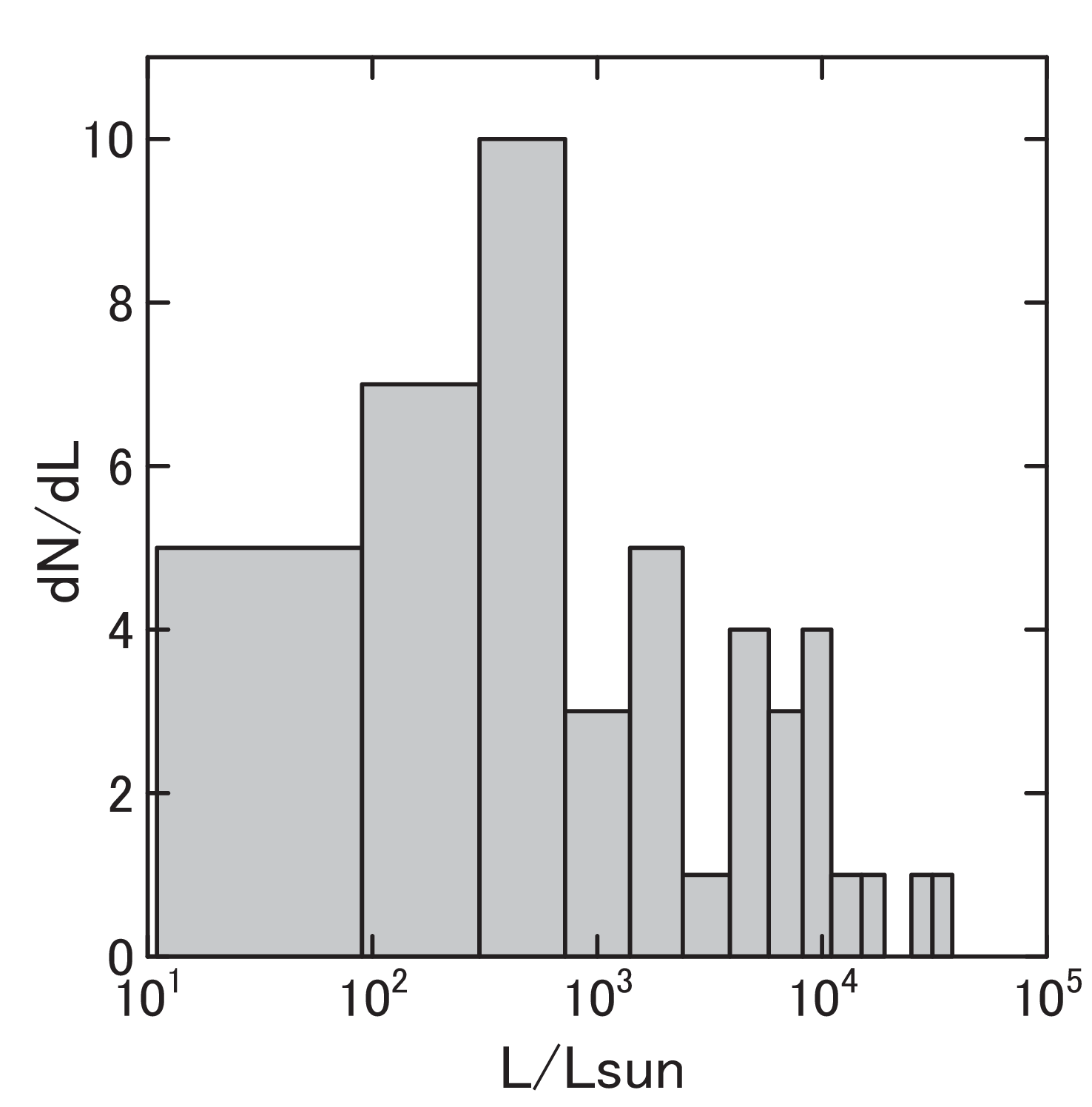}  
\end{center}
\caption{Frequencies of $\Rbow$, PA offset from vector perpendicular to local arm (dashed line in figure \ref{zoo}), and estimated $\L$.  }
\label{histo}
	\end{figure}

 Figure \ref{histo} shows the frequency distributions of $\Rbow$, estimated $\L$, and offset of PA ($\delta$PV) from vertical direction to the local spiral arm (dashed line in figure \ref{zoo}). In the statistics, we removed GCH/MBS with $\Rbow>150$ pc, which are mostly weak dark lanes without luminous \Ha nebulae (dashed arcs in figure \ref{zoo}).

The PA offset is concentrated around $\delta PA \sim 0\deg$, indicating that the bows (cone) develop perpendicularly to the arms toward the outer inter-arm region. The cones are open toward the down-stream side of the gas flow with the galactic rotation velocity with respect to the spiral density wave at rigidly rotating at the pattern speed slower than the galactic rotation.

	\begin{table} 
\caption{Bow head position (offset from nucleus), $\Rbow$, PA of cone axis, and $\L$.}
\begin{tabular}{llllll} 
\hline 
\hline 
 & $\delta$RA(")&$\delta$Dec(")&$\Rzero$(pc)& PA(deg)&$\L(\Lsun)$\\  
 \hline  A & 147.7&     -54.0&     160.4&     -60.1&  0.47E+05\\
     & 139.7&     -47.3&      71.3&      45.2&  0.41E+04\\
     & 137.7&     -51.0&     177.1&      82.8&  0.63E+05\\
     & 143.5&     -38.4&     169.6&      67.6&  0.55E+05\\
     & 135.9&     -29.8&      46.8&       0.0&  0.12E+04\\
     & 133.5&     -26.6&      52.5&      72.7&  0.16E+04\\
     & 133.3&     -17.4&      47.2&      -5.5&  0.12E+04\\
     & 143.1&     -13.3&      83.9&      -0.8&  0.67E+04\\
\hline  B & 126.4&      20.7&      39.4&       3.5&  0.69E+03\\
     & 134.2&      24.6&      91.6&      54.9&  0.87E+04\\
     & 140.0&      24.4&      34.9&      43.4&  0.48E+03\\
     & 149.9&      31.1&      31.4&      73.1&  0.35E+03\\
     & 164.9&      28.2&      61.2&     -72.9&  0.26E+04\\
     & 152.7&      35.0&      27.4&      43.7&  0.23E+03\\
     & 157.0&      40.5&      39.0&      61.2&  0.67E+03\\
     & 159.6&      44.9&      25.9&      32.2&  0.20E+03\\
     & 160.7&      48.8&      35.5&      52.8&  0.50E+03\\
     & 163.8&      52.0&      23.8&      38.3&  0.15E+03\\
     & 131.7&      44.5&      25.5&     -21.0&  0.19E+03\\
     & 126.6&      59.5&      16.6&      43.6&  0.51E+02\\
     & 131.3&      30.6&      51.4&      71.5&  0.15E+04\\
     & 139.2&      32.9&     104.4&      48.2&  0.13E+05\\
     & 142.1&      34.4&      79.2&      33.2&  0.56E+04\\
     & 141.0&      40.4&     142.5&      55.2&  0.33E+05\\
     & 148.4&      38.7&      15.6&      22.7&  0.43E+02\\
     & 153.7&      42.1&      20.4&      52.2&  0.95E+02\\
     & 155.0&      44.2&      19.6&      51.9&  0.85E+02\\
     & 149.8&      48.7&     134.2&      45.2&  0.27E+05\\
     & 153.9&      54.3&      83.5&      48.6&  0.66E+04\\
     & 156.6&      54.1&     191.0&      24.2&  0.78E+05\\
     & 144.0&      58.5&     118.6&      -5.3&  0.19E+05\\
\hline  C  & 70.2&      55.4&      37.7&      -4.4&  0.61E+03\\
     &  73.7&      53.9&      28.0&     -37.1&  0.25E+03\\
     &  82.7&      54.1&      84.7&     -31.6&  0.68E+04\\
     &  98.1&      50.5&      54.2&     -34.0&  0.18E+04\\
     &  90.0&      64.3&      94.8&     -35.8&  0.96E+04\\
     & 109.6&      58.9&      45.6&     -47.0&  0.11E+04\\
     & 106.2&      63.5&      91.7&     -22.4&  0.87E+04\\
\hline  D & 20.9&      25.8&      37.1&     -40.5&  0.57E+03\\
     &  26.8&      23.5&      17.3&     -47.1&  0.58E+02\\
     &  34.7&       9.8&      19.9&     -43.3&  0.88E+02\\
     &  36.3&       5.8&      33.4&     -46.1&  0.42E+03\\
\hline  E & 29.6&     -58.9&      59.2&      64.3&  0.23E+04\\
     &  33.2&     -49.9&      90.3&      58.8&  0.83E+04\\
     &  28.4&     -47.8&      32.8&      40.4&  0.40E+03\\
     &  34.9&     -40.3&      38.3&      62.2&  0.63E+03\\
     &  32.8&     -34.8&      24.3&      76.0&  0.16E+03\\
     &  36.8&     -29.2&      79.6&     -82.0&  0.57E+04\\
     &  43.4&     -29.8&      52.5&     -35.4&  0.16E+04\\
     &  42.8&     -22.1&      76.0&     -21.4&  0.49E+04\\
\hline 
\end{tabular} 
\label{tabbow} 
	\end{table} 
         
\section{Qualitative Models}

\subsection{Bow shock}
 .

\begin{figure} 
\begin{center}        
\includegraphics[width=6cm]{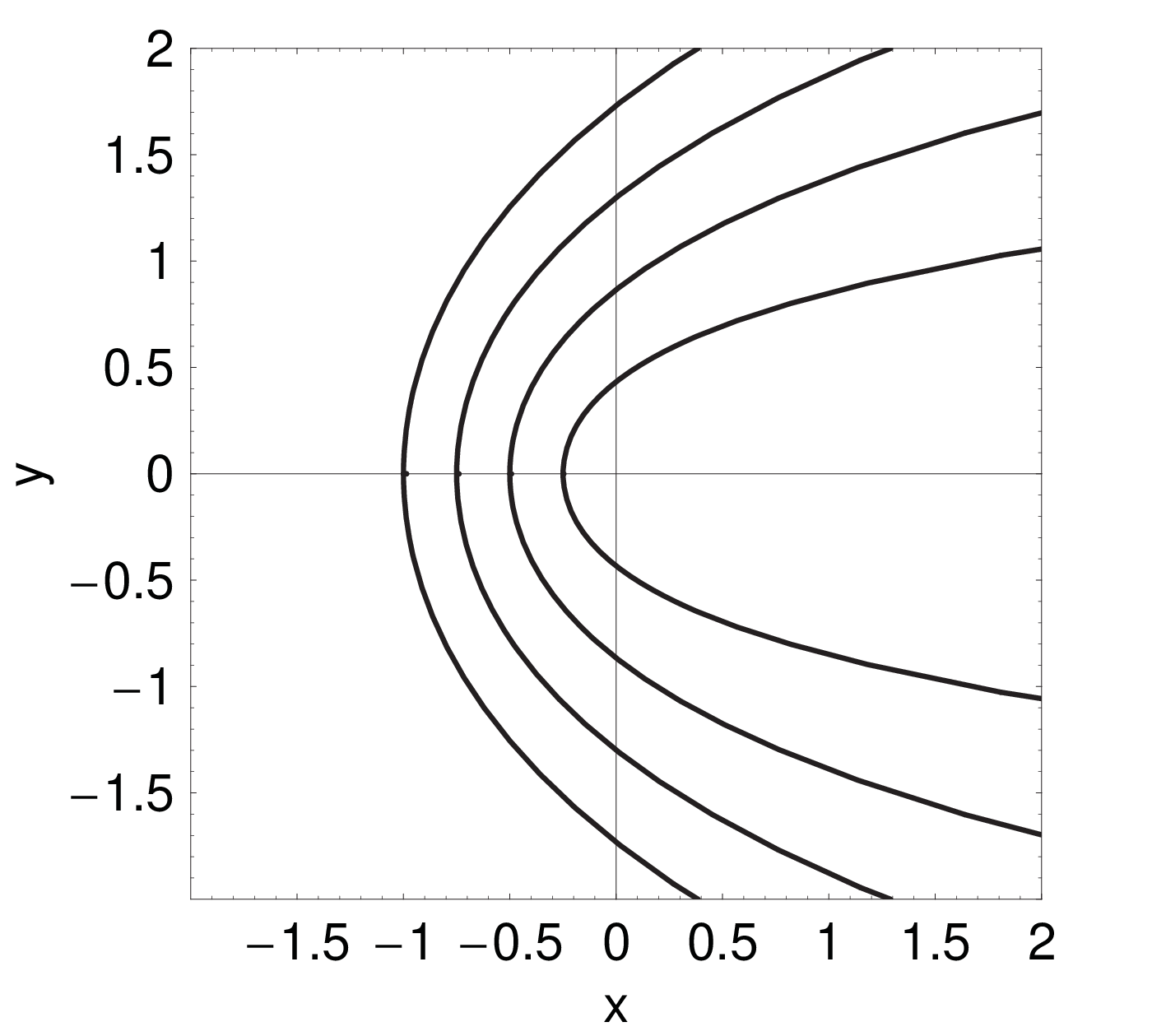}   \\
\includegraphics[width=5cm]{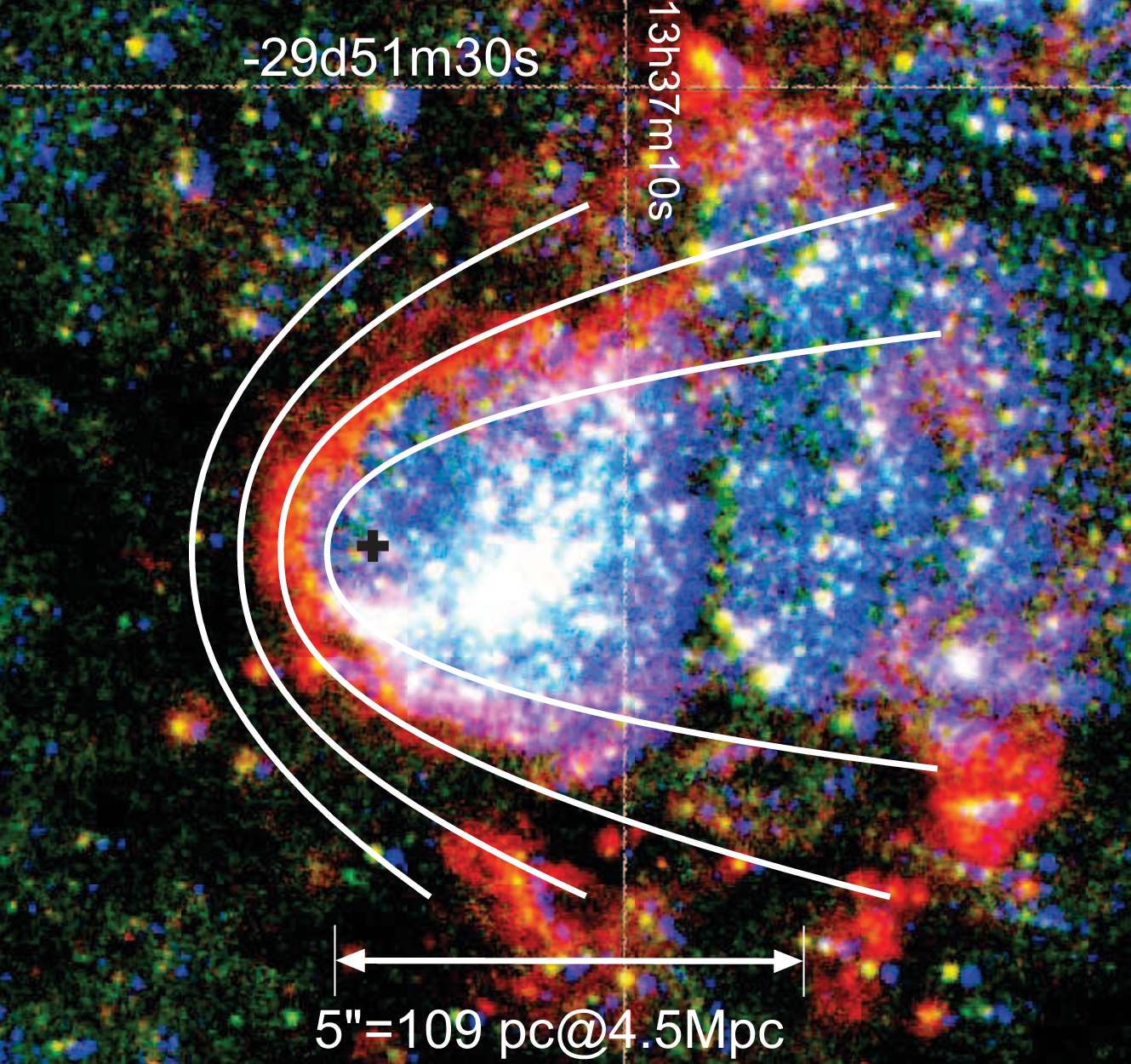}    
\end{center}
\caption{Calculated bow shock front (top), and the same overlaid on a Type III GCH in M83 from figure \ref{zoo} with arbitrary scaling (bottom). }
\label{bowshock}
\begin{center}     
\includegraphics[width=6cm]{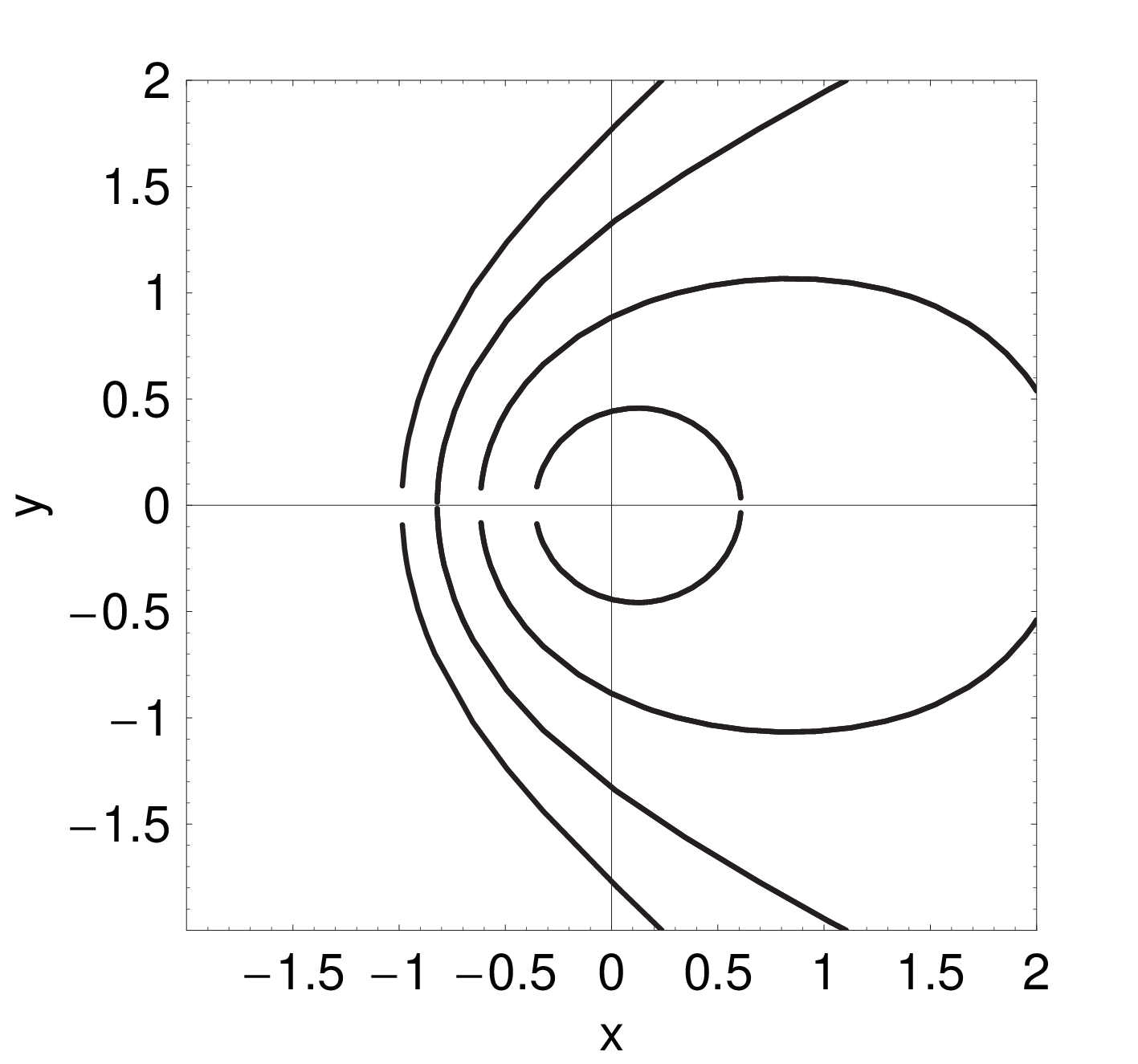} \\
\includegraphics[width=5cm]{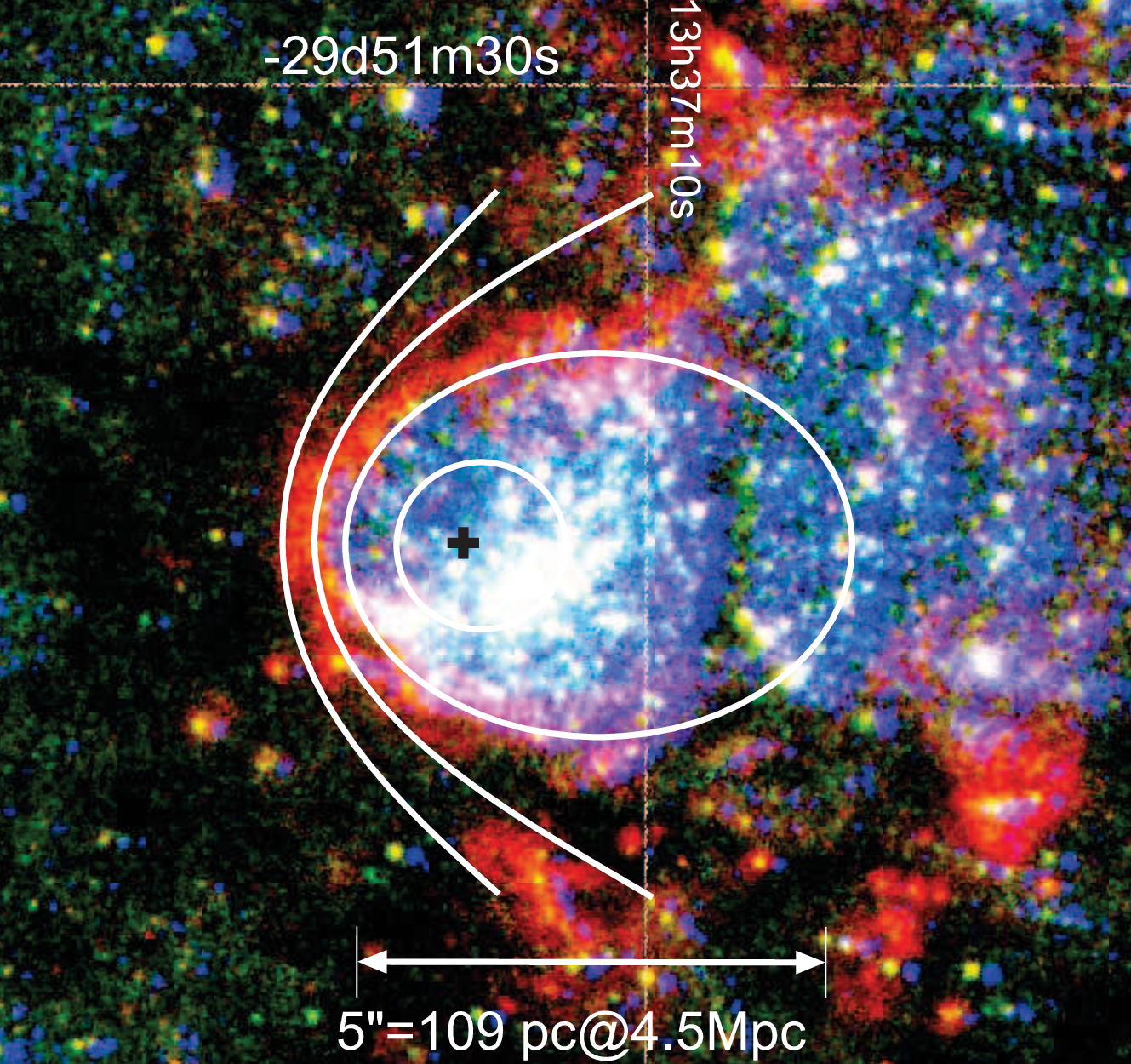}       
\end{center}
\caption{Cometary front of an HII region for different central luminosities, and the same overlaid on a GCH in M83 by arbitrary scaling.}
\label{comHII}
	\end{figure}

In our recent paper (Sofue et al. 2018) we modeled the MBS G30.5 by applying the Wilkin's (1996) analytical model for stellar-wind bow shock (figure \ref{G30}). The distance $Q$ of a bow front from the wind source is related to the elevation angle $\phi$ through
\be
Q(\phi)=\Rbow\ {\rm cosec}\ \phi \sqrt{3(1-\phi\ {\rm cot}\ \phi}).
\ee 
Here, $\Rbow$ is the stand-off radius defined as the distance of the front on the galactic plane from the wind source, which is measured as the smallest curvature of the bow head facing the gas flow. It is related to the momentum injection rate $\dot{m}_{\rm w}$ by wind of velocity $V_{\rm w}$ from the central star and ram pressure by the in-flowing gas from outside by
\be
\Rbow=\sqrt{\dot{m}_{\rm w} V_{\rm w} \over 4 \pi \rho V^2}.
\label{eq_bshock}
\ee 
In case of G30.5  we obtained $\Rbow\sim 54$ pc (Sofue et al. 2018), as shown in figure \ref{G30}, where bow shapes calculated for $\Rbow=25$, 50 and 75 pc are shown. Figure \ref{bowshock} shows the same calculated result compared with a GCH observed in region C of M83.

\subsection{HII cone}

In order to model the front shape of an HII region in inhomogeneous ISM, we first consider the Str\"omgren radius in a uniform ISM given by 
\be
 \Rhii \simeq \left({3 \nuv \over 4 \pi \ni\ne\ar }\right)^{1/3} 
 \label{strsph}
 \ee 
where $\nuv$ is the UV photon number radiated by the OB stars and $\ar \sim 4\times10^{-13}  {\rm cm^{-3}s^{-1}}$ is the recombination rate, $\ni$ and $\ne$ are the ion and eletron densities, respectively. 

We, then, assume that this equation holds in each small solid angle at any direction in which the density is assumed to be constant. We neglect the dynamical motion of gas inside the HII region. This approximation gives a qualitative shape of the front.

The ISM density variation is represented by an exponentially decreasing function in the $x$ direction perpendicular to the spiral arm with scale width $x_0$.
The $z$-directional (vertical) density profile is assumed to be represented by an inverse hyperbolic cosine function with scale thickness $z_0$.
Thus, the density in the galactic shock wave is expressed by
\be
n=n_0 \ \[\exp \(-{x\/x_0}\)+\epsilon_1\] \[{\rm cosh}^{-1}\( {z\/z_0}\)+\epsilon_2\] .
\label{nxz}
\ee
Here, $n_0$ is a constant, and $\epsilon_1$ and $\epsilon_2$ are  constants representing the relative background densities in the galactic plane and halo, respectively. The density increases exponentially at negative $x$, but decreases to the inter-arm value of $\epsilon_1$ at  $x\le -2 x_0$, beyond which the front cannot reach in the present cases.

The ion and electron densities are assumed to be related to the neutral gas density $n$ through
\be
\ne\sim \ni \sim n {\Tn \/\Te},
\label{ne}
\ee
where $T_{\rm n}$ and $T_{\rm e}$ are the temperatures of neutral and HII gas.

We now define the representative radius $\Rhii$ as the equilibrium radius of a spherical HII region in uniform gas with density $n$ by
\be 
\Rhii \simeq \left[{3 \nuv \over 4 \pi \ar n^2 } \left(\Te \/\Tn \right)^2\right]^{1/3}.
\label{Rhiineu}
\ee
Rewriting $\nuv \sim \L/h\nu$ with L and $h\nu$ being the luminosity of the central OB stars and UV photon energy over $h 912$ A, respectively, we have
\be 
\Rhii \sim 
96.1 \({\L\/10^4} \)^{1\/3} \({n\/10^2}\)^{-{2\/3}} \({\Tn\/20}\)^{-{2\/3}} \({\Te\/10^4}\)^{2\/3} [{\rm pc}], 
\label{Rhiipc}
\ee
where $\L$ is measured in $\Lsun$,  $n$ in H cm$^{-3}$, and $\Te$  in K. 

Writing $\Rhii =\sqrt{x^2+y^2+z^2}$, we can express  $y$ as a function of $x$ and $z$, which represents the front shape of a GCH as
\be
y=\sqrt{r_0^2 \(e^{-{x\/x_0}}+\epsilon_1\)^{-4\/3} \(\cosh^{-1}{z\/z_0}+\epsilon_2\)^{{-4\/3}}-x^2-z^2}.
\label{yxz}
\ee
Here, $r_0$ is a parameter depending on the luminosity of the central OB association counted in terms of the head curvature of GCH $\Rzero$, which represents the radius for the neutral gas density $n=n_0$ at $(x,y,z)=(0,0,0)$.
If we measure $\Rzero$, which is assumed to be equal to $\Rbow$, we can estimate the UV luminosity $\L$ of the exciting OB stars for a fixed gas density $n_0$.
        
In figure \ref{comHII} we show calculated HII fronts for $r_0=0.5$, 0.75, and 1 in the $(x,y)$ plane calculated for $\epsilon_1=0.1$. The scales are normalized by the scale length $x_0$. The bottom figure is an overlay of the caculated front on a GCH in M83, where the scale is adjusted arbitrarily to fit the image.

\begin{figure} 
\begin{center}     
\includegraphics[width=5cm]{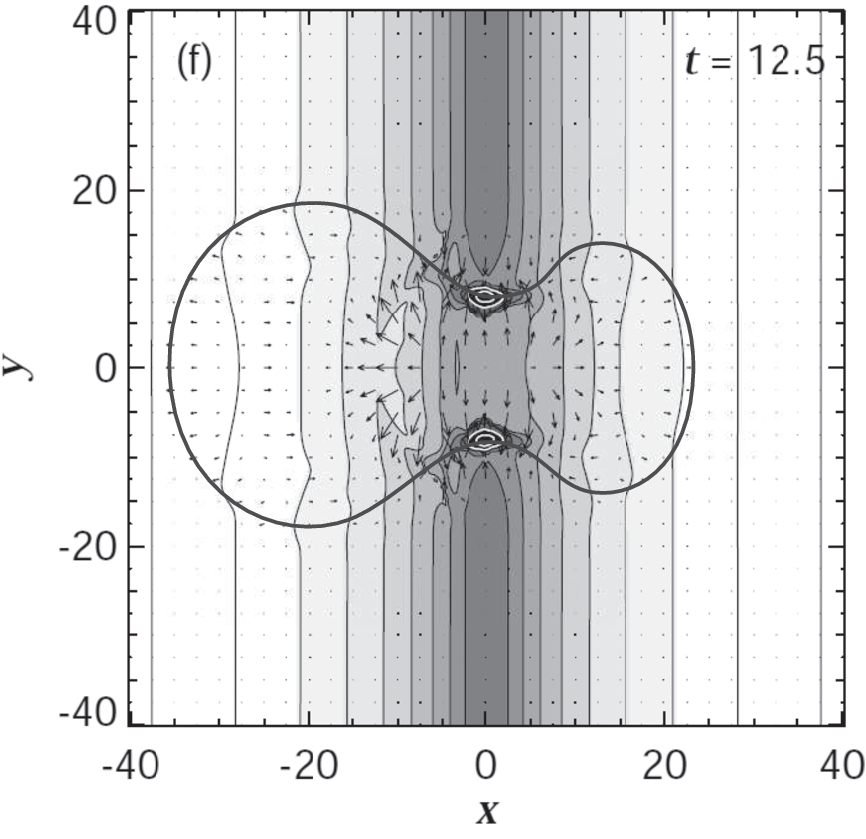}
\includegraphics[width=5cm]{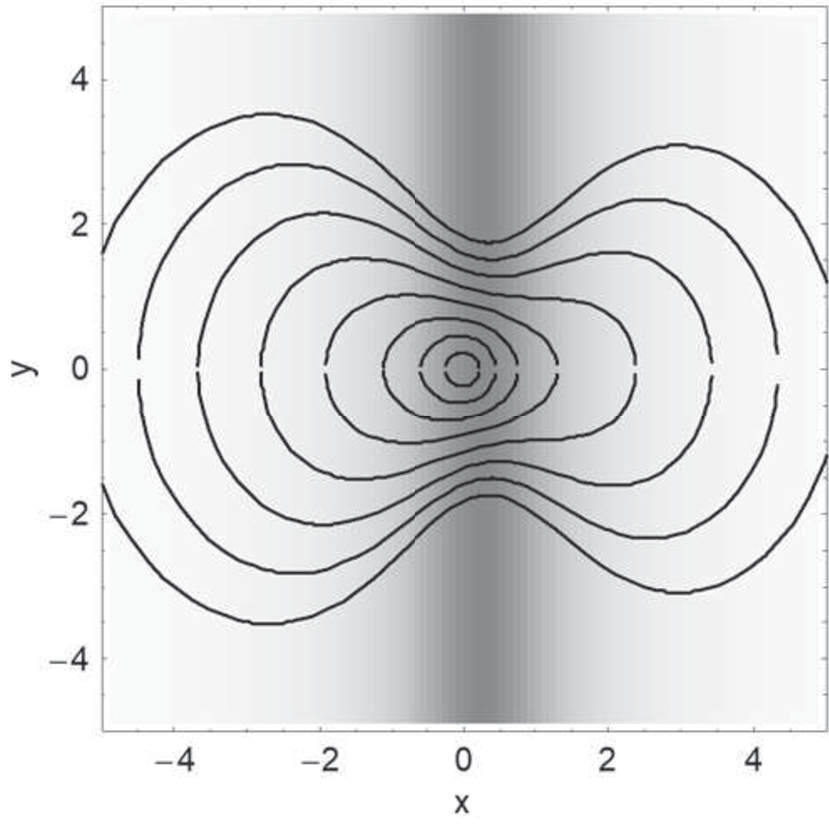}      
\end{center}
\caption{Comparison of hydrodynamical simulation by Fukuda and Hanawa (2000) of an off-center expansion of HII region (left) and front shape given by the present qualitative model (right) in a gas layer with density profile $\propto (1+x/x_0)^{-2}$.}
\label{hanawa}
	\end{figure}

In order to confirm that the qualitative model adopted here can reasonably reproduce a result by hydrodynamical simulation, we compare a calculated result of the HII front shape expanded in a sheet with density profile $\propto (1+x/x_0)^2$ with a hydrodynamical simulation by Fukuda and Hanawa (2000). Figure \ref{hanawa} shows the comparison, and we may consider that the model is reasonable in so far as we are interested in qualitative analysis of the front shapes. 

\begin{figure} 
\begin{center}    
Type I\includegraphics[width=5.8cm]{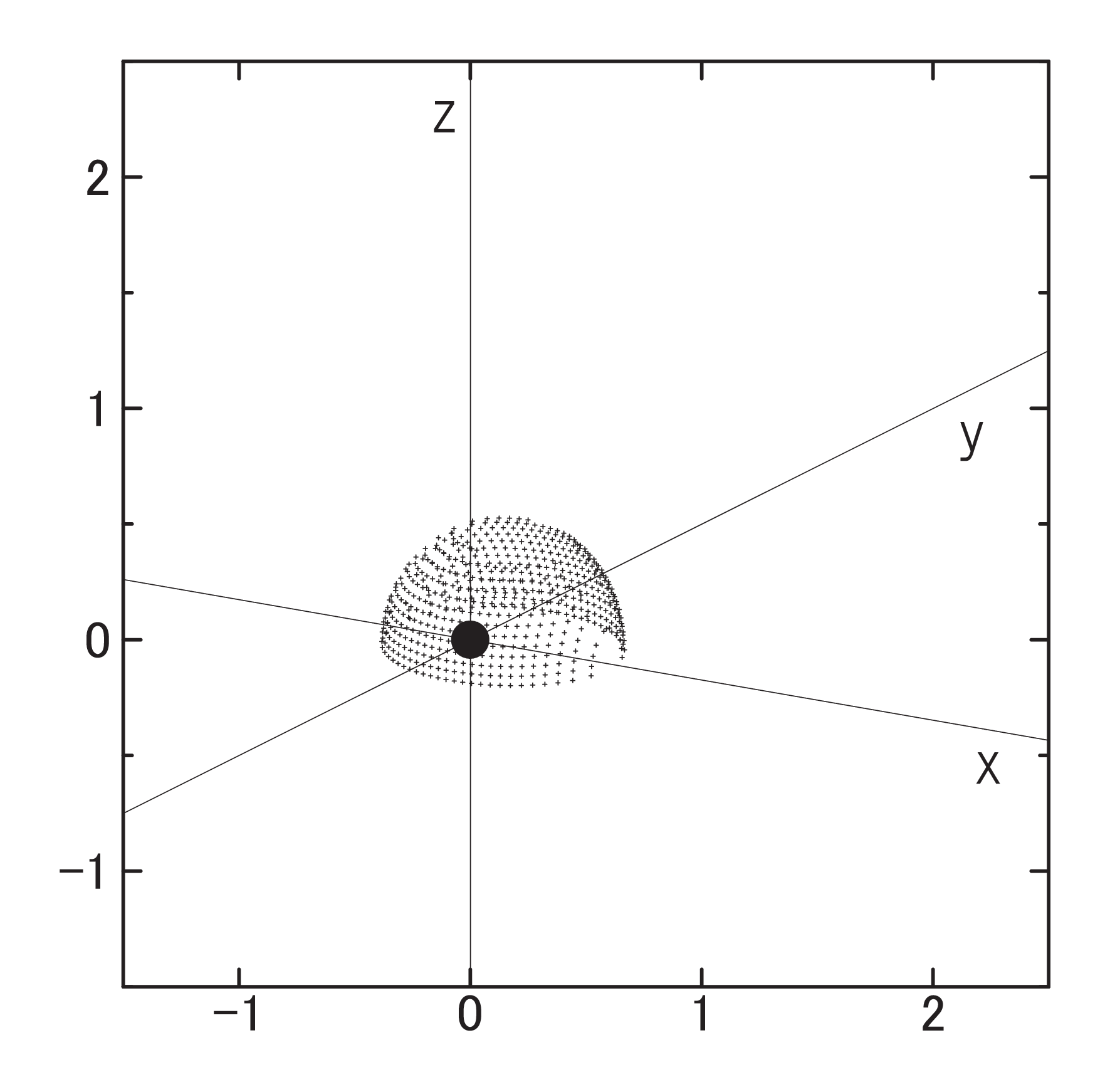}\\
Type II\includegraphics[width=5.8cm]{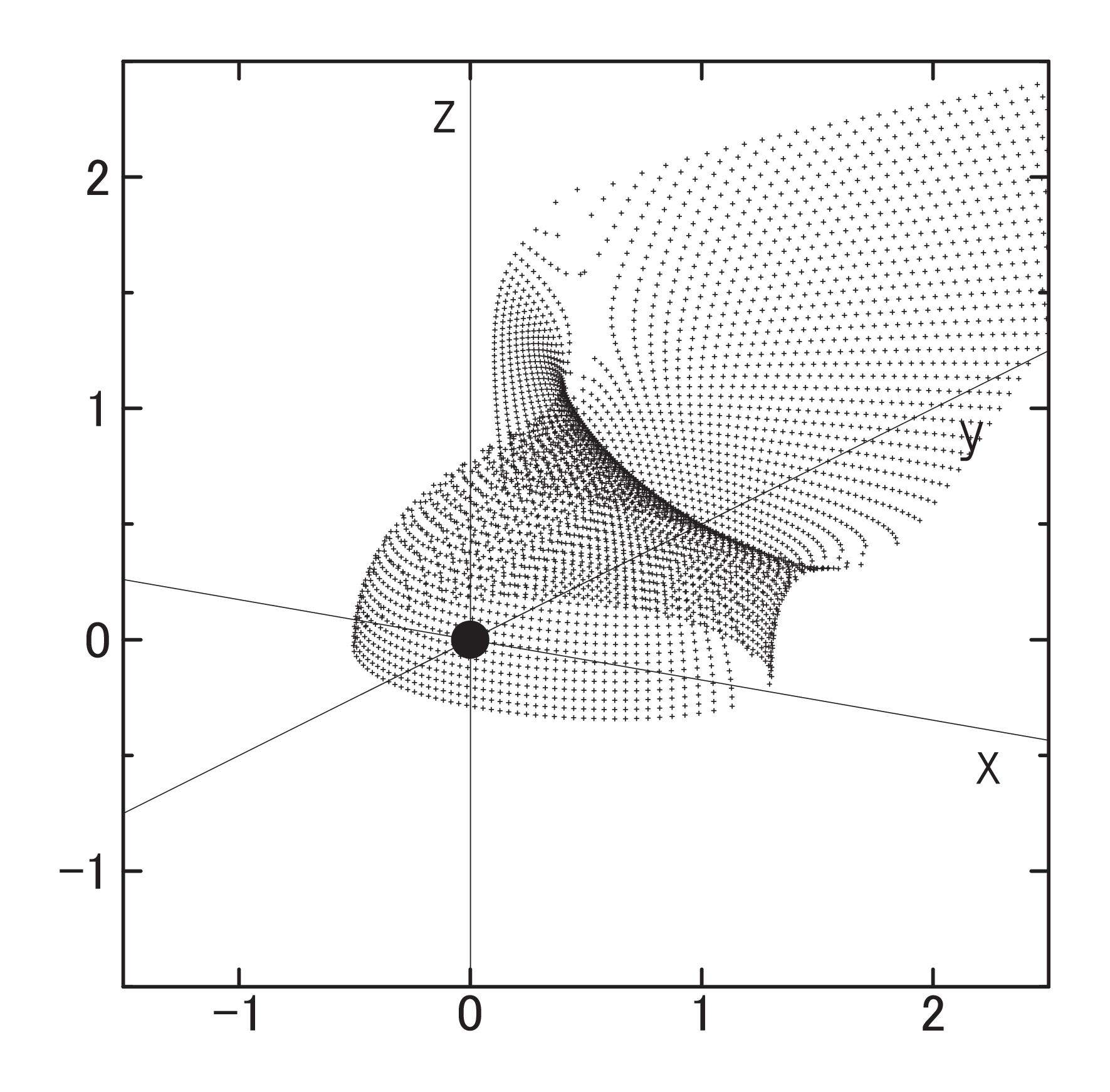}\\
Type III\includegraphics[width=5.8cm]{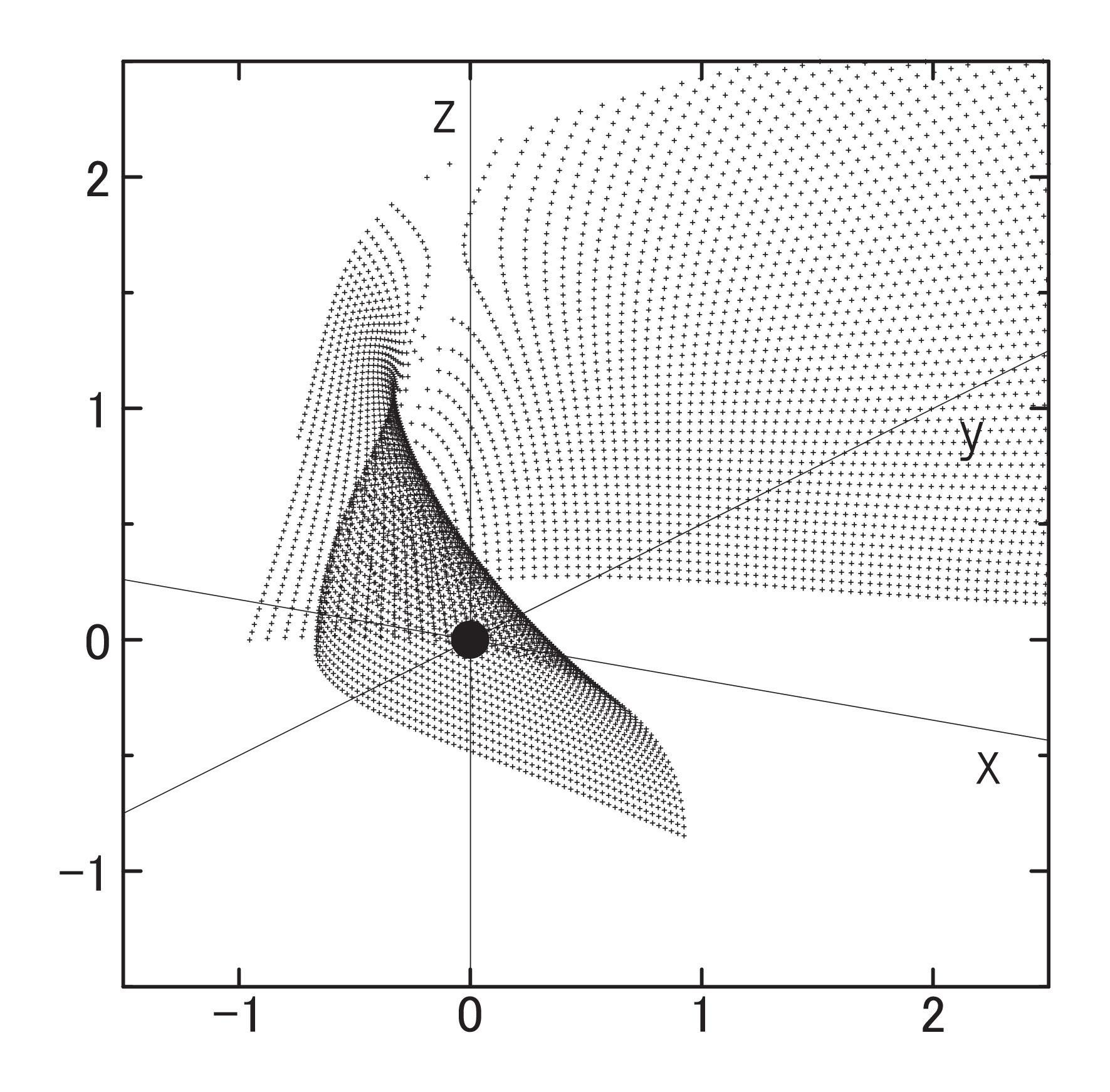}  
\vskip 0.7cm 
\end{center}
\caption{HII region shapes of Types I (bubble). II (cylinder) and III (GCH cone) according to $r_0=0.5$, 0.7, and 1 for fixed parameters of $x_0=1$, $z_0=1$, $\epsilon_1=0.1$ and $\epsilon_2=0.01$.  }
\label{model_3d} 
	\end{figure}

Figure \ref{model_3d} shows 3D front shapes for $r_0=0.5$, 0.7, 1 and 1.5 for $x_0=1$, $z_0=1$, $\epsilon_1=0.1$ and $\epsilon_2=0.01$, where the scales are normalized by $\Rhii$.
The front shapes vary according to the parameter $r_0\propto \Rzero \propto L^{1/3}n^{-2/3}$. 
For low luminosity OB clusters, the HII front shows an spherical shape of Type I. As the luminosity increases, the front is elongated in the $z$ direction, resulting in Type II bipolar cylinder open to the halo. As $r_0$ increases further, the shape becomes more open into the inter arm region and to halo, resulting in Type III for GCH.  
        
	\begin{table}  
\caption{representative GCH parameters}
\begin{tabular}{ll}  
\hline 
\hline 
Parameter & Value  \\   
\hline  
UV luminosity &$\L \sim 10^4\Lsun$   \\ 
Neutral gas density &$n \sim 10^2$ \Hcc \\
Temp. of neutral gas & $\Tn \sim 20$ K  \\
Temp. of HII gas &$\Te$ $\sim 10^4$ K \\ 
Scaling HII radius & $\Rzero \sim 96.1$ pc \\

\hline
Cone-head radius &$r_0/\Rzero= 0.5,\ 0.75,\ 1$ \\
Shock scale length & $x_0/\Rzero=1$ \\
Disk scale height &$z_0/\Rzero=1$ \\
Backgr. density in disk &$\epsilon_1=0.1$ \\
--- in halo &$\epsilon_2= 0.01$ \\

\hline  
\end{tabular}
\label{tabpara}  
	\end{table}    
 
\subsection{UV luminosity}

\co integrated intensity toward dark lanes in M83 has been observed to be $\Ico\sim 150$ \Kkms (Egusa et al. 2018), which yields molecular gas density of $n\sim 200$ \Hcc for an assumed disk thickness of $\sim 50$ pc and conversion factor of $2\times 10^{20}$ \Hsqcm (\Kkms)$^{-1}$ (Bolatto et al. 2013). Given the gas density, UV luminosities $\L$ of exciting OB stars are estimated for the measured $\Rzero\sim \Rbow$ using equation (\ref{Rhiipc}) as 
$\L \sim 4\times 10^2 (\Rzero/96.1)^3 \Lsun$. 
Thus estimated luminosities range from $\L\sim 10^3$ to $\sim 10^5\Lsun$ (table \ref{tabbow}), and may be compared with the visual luminosities of $L_{\rm v}\sim \times 10^4 - 10^6\Lsun$ (absolute visual magnitudes $M_{\rm v}\sim -6$ to -10) of OB clusters in M83 (Chander et al. 2010).

\subsection{GCH sheathed inside MBS}

MBS is composed of low-temperature and high-density molecular gas. On the other hand, GCH is high-temperature and low-density HII gas expanded around OB clusters. However, both MBS and GCH show a similar morphology (figures \ref{bowshock} and \ref{comHII}), and occur simultaneously. An HII region encountering a supersonic gas flow in galactic rotation makes a bowshock on its up-stream side, and is blown off toward the down-stream making open cone shape. 

The interaction of MBS and GCH may cause mutual deformation. The expansion of GCH is suppressed by MBS on the up-stream side, while it becomes more free in the tail. The side wall of MBS also suppresses the HII expansion, and guides the HII gas into a more collimated cone. 

\subsection{Dual-side compression and wavy sequential star formation}

The GCH and MBS act as a mutually triggering mechanism of star formation in such a way that GCH compresses the stacked molecular gas at the bow head from inside, and MBS compresses the molecular clouds from outside. The supersonic flow by the spiral arm's gravitational potential convey and supply molecular clouds to the bow head. 

Thus, the bow head becomes an efficient star forming site by the "dual dynamical compression" sandwitched by MBS and GCH. This enhances galactic-scale compression of molecular clouds along the galactic shock, and makes the star formation more efficient than usually considered such as due to cloud-cloud collisions (e.g. McKee and Ostriker 2007).

MBS and BCH appear in a chain along the arms (figures \ref{zoo} and \ref{zoo-II}), and compose a wavy array of SF regions along the galactic shock wave. Furthermore, the bow waves from MBS encounters the neighboring waves, triggering SF in dense regions between the MBS in a sequential way (figure \ref{illust}). 

\subsection{High SF rate}

It has been argued that cloud-cloud collision is an efficient mechanism of star formation (McKee and Ostriker 2007), and the star formation rate (SFR) is given by $1/\tc$, where $\tc$ is the cloud-cloud collision time given by
\be 
\tc \sim {\rhoc  \Rc\/\rhom \sigma}\sim 10^7\ {\rm y}.
\ee
Here, $\rhoc\sim 10^3$ \Hcc is GMC's gas density, $\rhom\sim 200$ \Hcc is the averaged density of molecular gas, $\Rc\sim 30$ pc is the mean cloud radius, and $\sigma\sim 7$ \kms is the ISM turbulent velocity. 

On the other hand the cloud-bow head collision time $\tb$ is given by
\be 
\tb \sim {\rhoc  \Rc \/\rhom (\V-\Vp)\sin\ p} \sim 10^6\ {\rm y},
\ee
where $\V\sim 200$ \kms is the rotation velocity of the galaxy, $\Vp\sim R \Op$  with $R\sim 2$ kpc and $\Op\sim 20$ \kms kpc$^{-1}$ is the pattern velocity, and $p\sim 30\deg-80\deg$ is pitch angle of the spiral arm or the gaseous bar in M83.

We may therefore conclude that the SF rate $1/\tb\sim 10^{-6}$ y$^{-1}$ due to the galactic shock wave assisted by the MBS+GCH dual compression is an order of magnitude higher than the SF rate $1/\tc \sim 10^{-7}$ y$^{-1}$ due to cloud-cloud collisions.  

\subsection{Cometary vs cylindrical types}

Our model calculation showed that the cometary structure is more efficiently produced when the density gradient is sharper toward the down-stream side of the arm  and flow velocity against the bow head is faster. Such condition may be present in grand-designed arms and bars with significantly faster galactic rotation than the pattern speed. In fact, cometary HII regions are generally observed in two-armed interacting spiral galaxy like M51 or in barred spirals like M83 and NGC 1300.

On the other hand, HII regions in flocculent arms such as M33 and NGC 2403 tend to show cylindrical or spherical morphology. This may be explained if the arms are co-rotating with the galactic rotation, that an HII region expands more symmetrically in the arm's center. Similarly, in distant arms in corotation with the galactic rotation, HII regions also tend to have spherical or cylindrical morphology. These facts may be used to measure the corotation radius by observing the directions of HII cones with respect to the arms.

In order to check this idea, we have traced round, or irregular MBS and GCH in the regions outside of the corotation in regions G to K in figure \ref{zoo-II}, where the big ellipse and long dashed lines indicate the corotation circle of radius $140''$ at position angle $225\deg$ and inclination $24\deg$ (Hirota et al. 2014). Although the tracing was rather inaccurate because of the lack in bright and clear edged HII regions, we found a number of irregular cometary HII regions and less clear MBS, as indicated by dashed arcs and small ellipses. The fraction of irregular cases are larger than in the inner arms as in figure \ref{zoo}. This may imply that the corotation indeed takes place near the indicated corotation circle.

\section{Summary}
 
Using optical images of nearby spiral galaxies taken with the HST (NASA APOD), we identified many sets of GCH and MBS in the barred spiral galaxy M83. 
We classified HII regions into Types I to III according to their degree of openness into the halo and disk. We argued that the GCH and MBS are general phenomena in galactic shock waves. 

The cone-shaped morphology of GCH is qualitatively explained by a model of an evolved HII sphere expanded in inhomogeneous ISM with steep density gradient, and the MBS is understood by the bow shock theory. Since in the actual galactic condition the GCH and MBS are coupled with each other, dual side compression of gas at the MBS/GCH heads makes the SFR more efficient by a factor of 10 than SFR by cloud-cloud collisions.
        
We have further examined high-resolution images of other galaxies from HST and Subaru Telescope, and found that GCH and MBS are general phenomena in grand-designed spiral arms and/or bars. A full atlas of GCH/MBS in nearby galaxies will be presented in a separate paper. 

\vskip 5mm
{\bf Aknowledgements} 
The optical images of M83 were reproduced from the web sites of STSci at http://www.stsci.edu/hst/wfc3/ and NASA at https://apod.nasa.gov/apod/.


\end{document}